\documentclass{article}
\usepackage{natbib}

\title{Synchrotron Self-Absorption Process in GRBs and the Isotropic Energy - Peak Energy Fundamental Relation}
\author{Piotrovich M. Yu., Gnedin Yu. N.\thanks{E-mail: gnedin@gao.spb.ru}, Natsvlishvili T. M.\\ Central Astronomical Observatory at Pulkovo,
Saint-Petersburg, Russia.}

\begin{document}

\maketitle

\begin{abstract}
The existence of strong correlation between the peak luminosity
(and/or bolometric energetics) of Gamma Ray Bursts (GRB) is one of
the most intrigue problem of GRB physics. This correlation is not
yet understood. Here we demonstrate that this correlation can be
explained in framework of synchrotron self-absorption (SSA)
mechanism of GRB prompt emission. We estimate the magnetic field
strength of the central engine at the level $B\sim 10^{14}
(10^3/\Gamma)^3 (1+z)^2$, where $\Gamma$ is the Lorentz factor of
fireball.

{\bf Key words:} gamma ray bursts, synchrotron radiation, magnetic
field.
\end{abstract}

\section{Introduction}

One of the recent important discoveries in gamma ray bursts (GRBs)
is the discovery that the peak luminosity $L_p$ (or/and the
bolometric energetics) of GRBs correlates with their peak energy
$E_p$ \citep{b1,b2,b7,b3}. This correlation allows to use it for
constraining the cosmological parameters (see review \citet{b4}).
\citet{b5} and \citet{b6} used the Amati relation for implications
to fireball models.

The Amati relation between the isotropic-equivalent time
integrated prompt energy $E_{iso}$ and the peak energy
$\acute{E}_p$ in the comoving rest frame takes a form:

\begin{equation}
\acute{E}_{iso}\sim E_p^2 (1+z)^2 \label{eq1}
\end{equation}

Recently \citet{b8} used the Amati relation as an example to test
the cosmological evolution of GRBs. He used the Amati relation in
logarithmic form (Amati, 2006):

\begin{equation}
\log{E_{iso}} = a + b \log{E_p}
\label{eq2}
\end{equation}

\noindent where $E_{iso}$ is the isotropic-equivalent energy
defined in the $1-10^4$ KeV band in the GRB frame. \citet{b2} has
obtained with a sample of 41 long GRBs the next values of
parameters in (2): $a = -3.35$, $b = 1.75$ ($E_p$ in KeV and
$E_{iso}$ in $10^{52}$ erg). These values have been obtained with
the least squares method. If one uses the maximum likelihood
method with an intrinsic dispersion the values of the numerical
parameters are $a = -4.04$ and $b = 2.04$ \citep{b2,b8}.

\citet{b7} have shown that also the peak luminosity $L_{p,iso}$ of
the prompt emission correlates with $E_p$ in the same way as
$E_{iso}$:

\begin{equation}
E_p\sim L_p^{1/2},\,\,(L_p\equiv L_{p,iso})
\label{eq3}
\end{equation}

The scatter appears to be similar to the scatter of the Amati
relation.

\citet{b3} found that after correcting isotopic energetics by the
collimation factor $E_{\gamma} = (1 - \cos{\theta}) E_{\gamma,
iso}$, where $\theta$ is the collimation angle, the collimation
corrected energy $E_{\gamma}$ also correlated with $E_p$. The
correlation is $E_p\sim E_{\gamma}^{0.7}$.

The next form of correlation has been obtained by \citet{b9}.
Their correlation is entirely phenomenological and involves three
observables and redshift. It is of the form $E_{iso}\sim E_p^2
t^{-1}$, where $t$ is the scaled characteristic time of the jet
development. If $t$ is close to unit, then the Liang and Zhang
correlation is entirely corresponding to the Ghirlanda
correlation.

At last, the Firmani correlation \citep{b10} links three
quantities of the prompt emission: the peak bolometric and
isometric luminosity $L_p$, the peak energy $E_p$ and
characteristic time $T_{0.45}$ which is time interval during that
the prompt emission is above a certain level. This correlation
takes a form: $L_p\sim E_p^{3/2} T_{0.45}^{-1/2}$, that is to be
similar to the Ghirlanda relation.

Above mentioned relations show that their base is really the
Yonetoku relation described by (3).

One of the interpretation of these relations has been recently
suggested by \citet{b11} and was later developed by \citet{b12}.
They showed that both the slopes and scatters in the Amati and
Ghirlanda relations, and the difference between them, are
quantitatively consistent with their model, in which the $E_{iso}
- E_p$ relation is due to viewing angle effect, i.e. consistent
with pure geometrical effect.

In this paper we would like to show that the Amati relation can be
received in framework of the synchrotron radiation mechanism with
self-absorption. It means that the central engine of powerful GRB
emitted in electromagnetic form $L_{iso}\sim 10^{52} - 10^{53}$
erg/s is tightly linked with a strong magnetic field. One cannot
exclude that such powerful central engine appears in a result of
the collapse of a star having substantial angular momentum. The
collapse of a star may be accompanied by the formation of a
quasi-static object - a spinar - whose equilibrium is maintained
by centrifugal forces and a strong magnetic field
\citep{b13,b14,b15}.

\section{Synchrotron Self-Absorption Process (SSA): Determination of Magnetic Field Strength}

The magnetic field strength of cosmological GRB is believed to be
estimated by assuming that the GRB energy spectrum is produced in
a result of synchrotron emission with the observed radiation peak
frequency $\nu_p$ at which the optical depth for synchrotron
self-absorption is equal to unity being known.

Then averaged magnetic field strength is derived by the relation
\citep{b16,b17}:

\begin{equation}
B = 10^{-5} b(\alpha)\left(\frac{\nu_p}{1 GHz}\right)^5
\left(\frac{\phi}{0,"001}\right)^4 \left(\frac{1 Jy}{S_p}\right)^2
\frac{\delta}{1+z} \label{eq4}
\end{equation}

The angular diameter - redshift relationship is derived as:

\begin{equation}
\phi = \frac{l (1+z)^2}{D_L} \label{eq5}
\end{equation}

\noindent where $\phi$ is an apparent angular diameter of the
radiative source, $l$ is the linear diameter of spherical source,
and $D_L$ is the luminosity distance, $S_p$ is the peak flux
density, $\delta$ is the boosting factor, the coefficient
$b(\alpha)$ depends on the index of synchrotron power law
spectrum. Its value lies in the region $b(\alpha)=2\div 3$, for
the wide range of values $\alpha$.

In our case one needs to take into account the effect of
collimation and jet aperture. It means that the real linear
diameter of GRB fireball is $l = l_{\bot} = \theta R$, where $R$
is the jet length.

The peak flux density $S_p$ can be presented in the form:

\begin{equation}
S_p = \frac{L_p}{4\pi D_L^2 \nu_p (1+z)}
\label{eq6}
\end{equation}

\noindent where $L_p$ is total luminosity and $L_p = L_{iso}$.

The peak energy $E_p$ is equal to $E_p = h\nu_p$. The comoving
peak energy is to be $\acute{E}_p = E_p (1+z)$.

\section{The $L_{p,iso}$ - $E_p$ relation in framework of SSA process}

The maximum luminosity can be estimated in terms of the
dissipation of magnetic energy:

\begin{equation}
L_p = L_{iso}\sim B^2 l^2 c
\label{eq7}
\end{equation}

The magnetic field of the SSA process is derived by (4).

Using (4)-(6) we obtain:

\begin{equation}
L_p\sim \left[\left(\frac{\nu_p}{1GHz}\right)^7
(1+z)^9\right]^2l^{10} \delta^2 L_p^{-4} \label{eq8}
\end{equation}

Let us mention that the boosting factor $\delta$ is to be

\begin{equation}
\delta = \frac{1}{\gamma (1 - \beta\cos{\theta})}\approx
\frac{2}{\gamma \theta^2}
\label{eq9}
\end{equation}

\noindent where $\theta$ is a jet angle, $\gamma$ is the Lorentz
factor and $\beta = v / c$. The comoving peak energy
$\acute{E}_p\equiv h\nu_p (1+z)\sim \gamma$.

The next relation takes place:

\begin{equation}
\left(\frac{\nu_p}{1 GHz}\right)^7 (1+z)^9 \sim (\acute{E}_p)^7
(1+z)^2
\label{eq10}
\end{equation}

Then the relation (8) transforms into:

\begin{equation}
L_{iso}\sim (\acute{E}_p)^{12} (1+z)^4 \theta^6
\frac{1}{L_{iso}^4} \label{eq11}
\end{equation}

\noindent and

\begin{equation}
L_{iso}^5 \sim E_p^{12} (1+z)^{16} \theta^6
\label{eq12}
\end{equation}

According to \citet{b18} (see also \citet{b4}):

\begin{equation}
\theta\sim \left(\frac{t_j}{1+z}\right)^{3/8} \left(\frac{n
\eta_{\gamma}}{L_{iso}}\right)^{1/8}\sim \frac{1}{(1+z)^{3/8}}
\frac{1}{L_{iso}^{1/8}} \label{eq13}
\end{equation}

This relation is valid on the case of relativistic jet propagation
in a homogeneous circumburst medium, and what's more
$\eta_{\gamma}$ is the efficiency for converting the explosion
energy to gamma rays, $n$ is the density of the ambient medium and
$t_j$ is the time scale of jet evolution.

Using (12) and (13) one can get the next equation respect to
$L_{iso} \equiv L_p$:

\begin{equation}
L_p^5\sim E_p^{12} (1+z)^{55/4} \frac{1}{L_p^{3/4}}
\label{eq14}
\end{equation}

\noindent or

\begin{equation}
L_p^{23/4}\sim E_p^{12} (1+z)^{55/4}
\label{eq15}
\end{equation}

The solution of (14) and (15) gives:

\begin{equation}
L_p^{23/48}\sim E_p (1+z)^{55/48}
\label{eq16}
\end{equation}

It means that we really obtain the Yonetoku correlation in
framework of synchrotron self-absorption process, i.e. $E_p\sim
L_p^{0.5}$. As $E_{iso}\approx L_p t_j$, the last relation is
equivalent also to the Amati relation $E_p\sim E_{iso}^{0.5}$.

\section{The Ghirlanda correlation}

\citet{b3} corrected the isotropic energetics by the factor
$(1-\cos{\theta})$ and found that collimation corrected the Amati
relation. \citet{b19} found that the collimation clustered jet
aperture angles into a narrow distribution producing a reservoir
of explosion energy $E_{\gamma} = (1-\cos{\theta}) E_{iso}$. From
this relation for $\theta\ll 1$ the new correlation $E_p\sim
E_{\gamma}^{0.7}$ is followed, if one uses (12) for the
collimation angle $\theta$. In this case instead of our (16) one
can get the correlation

\begin{equation}
L_{\gamma}^{23/36}\sim E_p (1+z)^{55/48}
\label{eq17}
\end{equation}

\noindent that is really approximately $E_p\sim L_{\gamma}^{0.7}$

\section{The Estimation of Magnetic Field Strength of Central GRB Engine}

Eq.(4) allows us to estimate the magnetic field strength of GRB
central engine. We use for this estimation the results of spectral
analysis of Swift long GRBs mode by \citet{b20}. They presented
the best fit of correlation between $\acute{E}_p$ and $E_{iso}$ in
the following logarithmic quantities:

\begin{equation}
\log{\frac{\acute{E}_p}{1 KeV}} = (2.25\pm 0.01) + (0.54\pm 0.02)
\log{\frac{E_{iso}}{10^{52.44} erg}}
\label{eq18}
\end{equation}

Such correlation coincides with Amati relation (1) and was
confirmed for 29 long GRBs observed by Swift satellite.

Using the relation: $E_{iso} = L_p t = 10^2 L_p t_2$, (4)-(6) and
the following expression for the linear size of GRB central engine
$l = \theta c t$, one can obtain the following relation instead of
(4):

\begin{equation}
B = 1.2\times 10^9 \left(\frac{\acute{E}_p}{1 KeV}\right)^7
E_{iso,52}^{-2} (1+z)^2 t_2^6 \theta^4 \delta
\label{eq19}
\end{equation}

From (9) it is followed the next expression for the boosting
factor $\delta\sim \Gamma$, where $\Gamma$ is the Lorentz factor
of the GRB fireball. The typical relation between the collimation
angle $\theta$ and the GRB Lorentz factor $\Gamma$ is $\theta\sim
1/\Gamma$. Using also relation (18) we transform (19) into:

\begin{equation}
B = 1.2\times 10^{14} (E_{iso,52})^{1.78} t_2^6 \Gamma_3^{-3}
(1+z)^2 G
\label{eq20}
\end{equation}

The typical Lorentz factor of GRB fireball is $\Gamma \sim 10^3
\Gamma_3$ \citep{b21,b22}. The estimation (20) gives for $z > 1$
the magnetic field strength as $B\geq 10^{14}$ G. This value is
quite well corresponded to magnetar or/and spinar models of GRBs
\citep{b15}. If our suggestion on magnetar model is valid, it
means that in our Galaxy four GRB events took place during its
evolution time.

\section{Summary}

We demonstrate that the most important relation for physics of GRB
between the peak luminosity (and the bolometric energetics) and
their peak energy can be obtained in the framework of synchrotron
mechanism of GRB emission taking into account the effect of
self-absorption. In a result it is success to estimate the
magnetic field strength of a central engine of GRB. Its value
appears at the level of $> 10^{14}$ G.

\section*{Acknowledgments}

This work supported by the RFBR grant 07-02-00535, the Program of
the Presidium RAS "Origin and Evolution of Stars and Galaxies",
the program of the Department of Physical Sciences of RAS "The
Extended Structure...".

\end{document}